\documentclass[twocolumn,showpacs,prl]{revtex4}
\usepackage{amssymb}
\usepackage{graphicx}
\usepackage{dcolumn}
\usepackage{bm}
\usepackage{amsmath}

\begin{document}

\title{Hyperfine Coherence in the Presence of Spontaneous Photon Scattering}

\author{R.~Ozeri}
\author{C.~Langer}
\author{J.~D.~Jost}
\author{B.~L.~DeMarco}
\altaffiliation[Present address: ]{Department of Physics,
University of Illinois at Urbana-Champaign, Urbana, IL 61801-3080}
\author{A.~Ben-Kish}
\altaffiliation[Present address: ]{Department of Physics,
Technion, Haifa 32000, Israel.}
\author{B.~R.~Blakestad}
\author{J.~Britton}
\author{J.~Chiaverini}
\author{W.~M.~Itano}
\author{D.~Hume}
\author{D.~Leibfried}
\author{T.~Rosenband}
\author{P.~Schmidt}
\author{D.~J.~Wineland}
\affiliation{NIST Boulder, Time and Frequency division, Boulder
Colorado 80305}


\begin{abstract}
The coherence of a hyperfine-state superposition of a trapped
$^{9}$Be$^+$ ion in the presence of off-resonant light is
experimentally studied. It is shown that Rayleigh elastic
scattering of photons that does not change state populations also
does not affect coherence. Coherence times exceeding the average
scattering time of $19$ photons are observed. This result implies
that, with sufficient control over its parameters, laser light can
be used to manipulate hyperfine-state superpositions with very
little decoherence.
\end{abstract}

\maketitle

Superpositions of hyperfine states of atoms have been the subject
of considerable experimental interest.  A good example is the role
they have played in the development of atomic clocks over the last
five decades \cite{bauch02}.  More recently, hyperfine coherences
of quantum-degenerate gases have been used to reveal their
intrinsic properties \cite{Cornell_02,Ketterle_03}. Atomic
hyperfine-state superpositions are also being investigated as
possible information carriers for quantum information processing
\cite{icap02}.

In many such experiments, laser light is used to coherently
manipulate the hyperfine superpositions with stimulated Raman
transitions. In addition, laser light can be used to trap atoms as
in the case of optical-dipole traps. Since light perturbs the
energies of hyperfine levels, imperfect control of laser-beam
parameters can lead to dephasing of the superpositions and loss of
coherence \cite{bible,Mechede_coherence,Mikkel04}.

Past experiments with neutral-atoms in dipole traps investigated
the coherence of hyperfine superpositions in the presence of light
\cite{Mechede_coherence,Mikkel04}. In these experiments the
dominant source of dephasing was noise in experimental parameters
such as fluctuations in the laser intensity or the ambient
magnetic field.

A more fundamental source of decoherence arises from spontaneous
scattering of photons \cite{Knight97}. Spontaneous scattering is
typically suppressed by detuning the laser frequency from allowed
optical transitions, but it cannot be eliminated completely.
Generally, if a spontaneously scattered photon carries information
about which hyperfine state scattered the light, the event
effectively measures the atomic state and the superposition
collapses. In contrast, if the scattered photon does not contain
this information then coherence is preserved. In this letter, we
verify this effect by means of an experimental study of the
hyperfine decoherence of a trapped $^9$Be$^+$ ion caused by
spontaneous scattering of photons from a non-resonant laser beam.
Our results show that coherence can be preserved in the presence
of spontaneous photon scattering.

Off-resonant spontaneous scattering is a two-photon process in
which the atom scatters a laser photon into an electromagnetic
vacuum mode. Following such a scattering event the atom can be
found in the same or a different internal state, corresponding to
elastic Rayleigh or inelastic Raman scattering, respectively. The
polarization and frequency of a Raman scattered photon depend on
the angular momentum and energy imparted to the atom and are
therefore entangled with the atomic internal state, as
demonstrated in \cite{Monroe1}. A single Raman scattering event
will therefore completely decohere a hyperfine superposition
\cite{Vogel_00}. In Rayleigh elastic scattering the atom's
internal states are not entangled with the scattered photon
\cite{Mollow1, Walther1}. Therefore, an atom initially prepared in
a hyperfine-state superposition will remain in this superposition
after the photon was scattered.

The scattering amplitude can be calculated by evaluating the
electric-dipole coupling between the initial ground-state and the
relevant excited state, and between the excited state and final
ground-state. When there are several excited states, the
scattering amplitude is given by a coherent sum over all
amplitudes of scattering through the different excited states.

\begin{figure}[tb]
\begin{center}
\includegraphics[width=8cm]{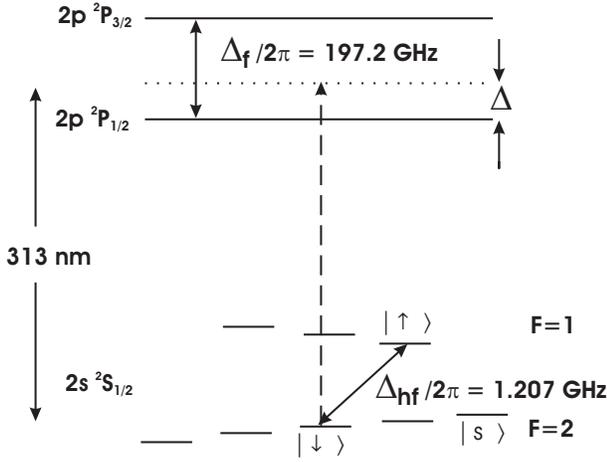}
\end{center}
\caption{The relevant energy levels of $^9$Be$^+$. The
ground-state $2s$ $^2S_{1/2}$ manifold consists of the $F=1$ and
$F=2$ hyperfine levels. Superposition states are composed of the
$|F=1, m_{F} = 1\rangle \equiv |\uparrow \rangle$ and $|F=2, m_{F}
= 0\rangle \equiv |\downarrow \rangle$ states. The excited-state
$2p$ $^2P$ manifold consists of the $J=1/2$ and $J=3/2$
fine-structure levels, separated by $\Delta_{f}/2\pi=197.2$ GHz.
The laser frequency, illustrated by the dashed arrow, is detuned
by $\Delta$ from the $|\downarrow \rangle$ to $^2P_{1/2}$
transition frequency.} \label{fig0}
\end{figure}

Specifically, the relevant energy levels in $^{9}$Be$^+$ are
illustrated in Fig. \ref{fig0}. Light is detuned from the $2s$
$^2S_{1/2} \rightarrow 2p$ $^2P_{1/2,3/2}$ transitions near $313$
nm. The $2p$ $^2P$ manifold has a fine structure that consists of
the $J=1/2$ and $J=3/2$ levels, separated by
$\Delta_{f}/2\pi=197.2$ GHz. The ion is illuminated with a laser
beam of intensity $I$. The laser polarization, $\hat{\sigma_k}$,
is characterized with respect to the magnetic field, which sets
the quantization axis for the ion, where $k = 0,+$ or $-$
correspond to parallel to the magnetic field ($\hat{\pi}$) or
right or left circular polarization, respectively. The rate of
photon scattering events in which an ion initially in state
$|i\rangle$ ends up in state $|f\rangle$ is given by the
Kramers-Heisenberg formula \cite{Loudon, cline94},
\begin{equation}
\Gamma_{i,f} = g^{2}\gamma|\frac{a^{(1/2)}_{i\rightarrow
f}}{\Delta} + \frac{a^{(3/2)}_{i\rightarrow f}}{\Delta -
\Delta_{f}}|^{2}. \label{rate}
\end{equation}
Here, $g=E\mu/2\hbar$, $E=\sqrt{2I/c\epsilon_{0}}$ is the laser
beam electric field amplitude, $c$ is the speed of light,
$\epsilon_{0}$ is the vacuum dielectric constant, and
$\mu=|\langle 2P_{3/2}, F=3,
m_{F}=3|\textbf{\emph{d}}\cdot\hat{\sigma}_{+}| 2S_{1/2}, F=2,
m_{F}=2\rangle|$, where $\textbf{\emph{d}}$ is the electric-dipole
operator. The effective amplitude $a^{(J)}_{i\rightarrow f} =
\sum_{q}\sum_{e \in J }\langle f
|\textbf{\emph{d}}\cdot\hat{\sigma}_{q}|e\rangle \langle e
|\textbf{\emph{d}}\cdot\hat{\sigma}_{k}| i \rangle/\mu^2$ is the
sum over amplitudes of scattering through all levels, $|e\rangle$,
in the $^2P_{J}$ manifold, $\Delta$ is the laser detuning from the
$^2S_{1/2} \rightarrow ^{2}P_{1/2}$ transition, and $\gamma/2\pi =
19.4$ MHz is the radiative lifetime of the excited states in the
$^2P$ manifold \cite{Remark1}.

The Raman scattering rate, $\Gamma_{Raman}$, is given by summing
over all the rates given by Eq. (\ref{rate}) where $i \neq f$. The
Rayleigh scattering rate, $\Gamma_{Rayleigh}$, is that when $i =
f$, and $\Gamma_{total}=\Gamma_{Raman}+\Gamma_{Rayleigh}$. The
matrix elements $a^{(1/2)}_{i\rightarrow f}$ and
$a^{(3/2)}_{i\rightarrow f}$ are identical in magnitude and
opposite in sign for Raman scattering, whereas they are equal in
sign for Rayleigh scattering. Therefore when $|\Delta|\gg
\Delta_f$ the two amplitudes in Eq. (\ref{rate}) destructively
interfere to suppress Raman relative to Rayleigh scattering. In
this limit the total scattering rate decreases as $1/\Delta^{2}$,
whereas Raman scattering alone scales as $1/\Delta^{4}$. This
suppression of population-changing Raman spontaneous scattering
has been observed previously by Cline {\em et al.}\
\cite{cline94}.

In our experiment, a single trapped $^{9}$Be$^+$ ion in a
superposition of two hyperfine states is illuminated by
off-resonant laser light. The coherence and population relaxation
rates are measured. The population relaxation rate provides the
Raman photon scattering rate, which is then compared to the
decoherence rate for different laser detunings.

We encode the superposition into the $|F=1, m_{F} = 1\rangle
\equiv |\uparrow \rangle$ and $|F=2, m_{F} = 0\rangle \equiv
|\downarrow \rangle$ ground-states. At a magnetic field of 0.01194
T, the two levels are separated by $\Delta_{hf}/2\pi = 1.207$ GHz.
At this field the energy difference between these two states does
not depend to first order on changes in the magnetic field, and
hence decoherence due to ambient magnetic field fluctuations is
negligible.

The ion is confined in a linear Paul trap, and its motional and
internal states are initialized by Doppler cooling and optical
pumping into the $|^2S_{1/2}, F=2, m_{F}=2\rangle\equiv |s
\rangle$, ``stretched" state. Raman transitions between different
hyperfine states are driven by a pair of co-propagating Raman
beams, detuned by $\Delta/2\pi\simeq82$ GHz, and separated by the
relevant hyperfine transition frequency. A resonant Raman pulse
transfers the ion population from $|s\rangle$ to
$|\uparrow\rangle$. Further Raman pulses are applied to manipulate
the superposition between the $|\uparrow\rangle$ and
$|\downarrow\rangle$ states. In a Bloch sphere representation
\cite{AllenEberly}, the rotation angle of a given pulse, $\theta$,
can be varied by changing the pulse duration. A $\pi$-pulse is
typically achieved with a duration of $\simeq 5$ $\mu$s. We
measure the coherence of the state ${\textstyle
\frac{1}{\sqrt{2}}}(| \downarrow \rangle + | \uparrow \rangle)$
using the Ramsey method of separated fields
\cite{Ramsey_NobelLecture}. This state is created with a
$\pi/2$-pulse that is applied to $|\uparrow \rangle$. After a
certain duration, a second $\pi/2$-pulse is applied with a phase
$\phi_R$ relative to the first pulse, where $\phi_R$ can be
varied. The population in the $| \uparrow \rangle$ state is then
measured by transferring this population to the state $|s\rangle$
followed by state-dependent resonance fluorescence \cite{bible}.
During a $200$ $\mu$S detection pulse, we typically detect $12$
photons if the ion is in the $|s\rangle$ state and approximately
one photon if it isn't.

To study decoherence, we illuminate the ion with a detuned,
$\hat{\sigma}_+$ polarized beam that is inserted between the two
Ramsey pulses. The decohering beam intensity has to be stabilized
($\lesssim0.1$ \%) to suppress Stark shift phase-noise
decoherence. In addition, a spin-echo sequence is implemented to
limit the bandwidth of remaining phase-noise to which the
superposition is susceptible \cite{AllenEberly, Mikkel04}.

\begin{figure}[tb]
\begin{center}
\includegraphics[width=8cm]{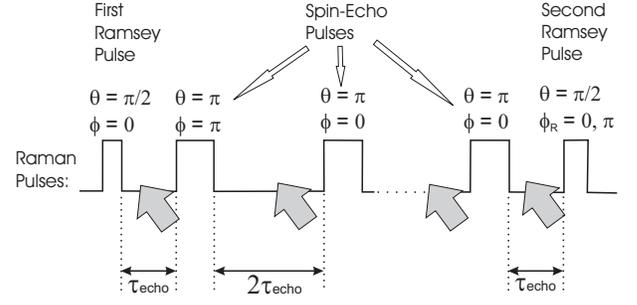}
\end{center}
\caption{Coherence relaxation pulse sequence. Two Ramsey pulses
surround a sequence of spin-echo $\pi$-pulses, of alternating
phase. The spin-echo pulses are separated by a time
$2\tau_{echo}$, during which the ion is illuminated by a
decohering beam, illustrated by the grey-filled arrows.}
\label{fig1}
\end{figure}

A typical experimental sequence is depicted in Fig. \ref{fig1}.
After a duration $\tau_{echo}$ following the first Ramsey pulse, a
sequence of $\pi$-pulses, separated by $2\tau_{echo}$ is applied.
The phase of subsequent $\pi$-pulses alternates between $0$ and
$\pi$ in order to correct for inaccuracies in the rotation angle.
Between the spin-echo pulses the ion is illuminated by the
decohering beam. The number of $\pi$-pulses is determined by the
maximum allowable value of $\tau_{echo}$  ($\simeq10$ ms), and
varies for different decohering beam detunings, between $2$ and
$18$. At a duration $\tau_{echo}$ after the final $\pi$-pulse, the
final Ramsey pulse is applied. The experiment is run twice, once
when $\phi_{R}$, the phase of the final Ramsey pulse equals $0$,
and once when it equals $\pi$. The population of the
$|\uparrow\rangle$ state is subsequently measured. These two
phases have been experimentally verified to provide the minimum
and maximum signals for a range of $\phi_R$, and therefore the
signal difference corresponds to the Ramsey contrast and
superposition coherence. The decrease of this difference as a
function of the decohering beam duration, $\tau$, is therefore a
measure of the decoherence rate, independent of population
relaxation.

\begin{figure}[tb]
\begin{center}
\includegraphics[width=8cm]{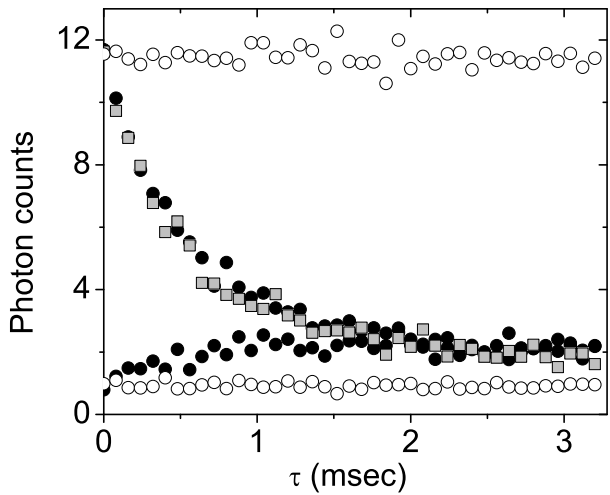}
\end{center}
\caption{Coherence and population relaxation data for a decohering
beam detuning of $\Delta/2\pi\simeq227.5$ GHz. The difference
between the two filled circles curves is proportional to the
superposition coherence at different times in the presence of the
decohering beam light. The difference between the two open circles
curves is proportional to the coherence in the absence of light.
The gray-filled squares are proportional to the $|\uparrow\rangle$
state population at different times. Very similar data was
recorded for the $|\downarrow\rangle$ state population. Every
point in the figure is the average photon count of 400
experiments.} \label{fig2}
\end{figure}

The filled circles in Fig. \ref{fig2} present the two measured
Ramsey signals vs. $\tau$ at a decohering beam detuning of
$\Delta/2\pi = 227.5$ GHz. The upper filled circles correspond to
the $\phi_{R}=\pi$ measurements, whereas lower filled circles are
the $\phi_{R}=0$ measurements. For $\tau\gtrsim 1$ ms, the two
traces are seen to collapse on top of each other. The two curves
do not collapse around their initial average value because
population relaxation happens on the same time scale as
decoherence. The decoherence time, $\tau_{dec}$, is found by an
exponential fit of the difference between the two curves.

The empty circles in Fig. \ref{fig2} show the data from an
identical experiment, except for the absence of the decohering
beam. As can be seen, no significant decoherence can be measured
during the measurement time in the absence of light.

We next measure the rates of population relaxation. The ion is
prepared in either the $|\uparrow\rangle$ or the
$|\downarrow\rangle$ state and illuminated by the decohering beam
light for a variable time $\tau$. The relevant state is then
transferred to $|s\rangle$ before detection. The gray-filled
squares in Fig. \ref{fig2} show the decay in signal when the ion
is prepared in the $|\uparrow\rangle$ state for $\Delta/2\pi =
227.5$ GHz. The decay of this signal is seen to agree with the
decay in the Ramsey, $\phi_{R}=\pi$, signal. Very similar curves
are measured when the ion is prepared in the $|\downarrow\rangle$
state.

For our initial conditions and laser polarization, the dynamics of
population relaxation of both the $|\uparrow\rangle$ and
$|\downarrow\rangle$ states can be modelled by the solution of two
coupled rate equations with independent loss rates,
\begin{equation}
P(t) = \frac{1}{2\alpha}e^{-\beta t}[\kappa \sinh(\alpha t) +
2\alpha \cosh(\alpha t)]. \label{Pop_relax}
\end{equation}
The constants $\alpha$, $\beta$, and $\kappa$ are different for
the $|\uparrow\rangle$ and $|\downarrow\rangle$ states, and, in
both cases, can be related to $\Gamma_{Raman}$. We extract
$\Gamma_{Raman}$ by fitting curves such as shown by the
gray-filled squares in Fig. \ref{fig2} to Eq. (\ref{Pop_relax}).

We normalize the measured rates by the measured Stark shift,
$\Delta_{St}$, in the $|\downarrow \rangle \rightarrow | \uparrow
\rangle$ transition frequency due to the decohering light, which
serves as an independent measurement of the laser intensity on the
ion. The Stark shift is measured by scanning the frequency of the
Ramsey pulses, and observing the shift of the Ramsey fringes with
the decohering beam applied in between the pulses.

\begin{figure}[tb]
\begin{center}
\includegraphics[width=8cm]{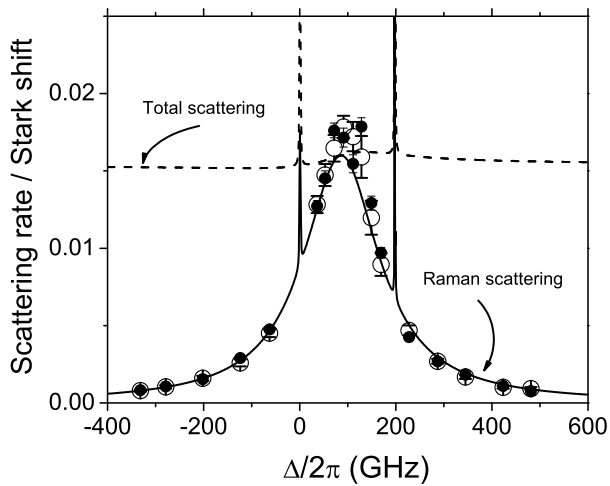}
\end{center}
\caption{Ratios of the different scattering rates and the
decoherence rate to the differential Stark shift for different
decohering beam detunings. The dashed line is the calculated
$\Gamma_{total}/\Delta_{St}$, which at large detunings is seen to
asymptotically reach a constant value of $\simeq 0.0967$ . The
solid line is the calculated $\Gamma_{Raman}/\Delta_{St}$. The
filled circles are the measured values of
$\Gamma_{Raman}/\Delta_{St}$ which are seen to be in reasonable
agreement with the calculated curve. The empty circles are the
measured values of $1/(\tau_{dec}\Delta_{St})$. The two sets of
data are seen to be in good agreement, thus demonstrating Raman
scattering of photons is the dominant causes of decoherence. The
suppression of the decoherence rate relative to the total
scattering rate is evident for large detunings.} \label{fig3}
\end{figure}

We calculate $\Delta_{St}$ by evaluating the difference in the
Stark shift of the two levels,
\begin{equation}
\Delta_{St} = g^{2}(\frac{a^{(1/2)}_{|\uparrow\rangle\rightarrow
|\uparrow\rangle}}{\Delta + \Delta_{hf}} +
\frac{a^{(3/2)}_{|\uparrow\rangle\rightarrow
|\uparrow\rangle}}{\Delta + \Delta_{hf} - \Delta_{f}} -
\frac{a^{(1/2)}_{|\downarrow\rangle\rightarrow
|\downarrow\rangle}}{\Delta} -
\frac{a^{(3/2)}_{|\downarrow\rangle\rightarrow
|\downarrow\rangle}}{\Delta - \Delta_{f}}). \label{Diff_stark}
\end{equation}
The amplitudes $a^{J}_{|\uparrow\rangle\rightarrow
|\uparrow\rangle}$ and $a^{J}_{|\downarrow\rangle\rightarrow
|\downarrow\rangle}$ are almost equal and the difference in Stark
shift is due primarily to the difference in detuning between the
two hyperfine-states. For $|\Delta|
>> \Delta_{hf}$, the differential Stark shift decreases according
to $1/\Delta^{2}$. The dashed line in Fig. \ref{fig3} shows
$\Gamma_{total}/\Delta_{St}$, the calculated total number of
photons which are scattered for one radian of Stark phase
evolution. This number asymptotically reaches a constant value of
$0.9579\times\gamma /\Delta_{hf} \simeq 0.0154$. As all of the
measured data were taken in the $|\Delta|
>> \Delta_{hf}$ limit the measured
$\Delta_{St}$ is a good measure of the total scattering rate,
almost independent of $\Delta$. The solid line shows the
calculated number of Raman scattered photons during the same
cycle, $\Gamma_{Raman}/\Delta_{St}$. This number decreases as
$1/\Delta^2$ for large laser detunings.

The filled circles in Fig. \ref{fig3} present the measured
$\Gamma_{Raman}/\Delta_{St}$ vs. the decohering beam detuning. The
measured data are seen to be in reasonable agreement with the
theoretical prediction (solid line).

The empty circles in Fig. \ref{fig3} show the measured ratio
between the decoherence rate and the differential Stark shift,
$1/(\tau_{dec}\Delta_{St})$ vs. the detuning. As can be seen the
measured points are in reasonable agreement with both the measured
and the calculated $\Gamma_{Raman}/\Delta_{St}$, and are well
below the $\Gamma_{total}/\Delta_{St}$ trace. For $\Delta/2\pi=
-331.8$ GHz more than $19$ photons are scattered on average before
coherence is lost.

Spin-changing suppression when $|\Delta| \gg \Delta_f$ also causes
the Rabi frequencies of stimulated Raman transitions to scale as
$1/\Delta^2$. The total number of photons that are scattered
during a $2\pi$-pulse asymptotically reaches a constant value
proportional to $2\pi\gamma /\Delta_f$ \cite{Wineland_London}.
However, since it is only Raman scattering that decoheres a
hyperfine-state superposition, infidelities due to spontaneous
scattering of photons will decrease as $1/\Delta^{2}$ for
$|\Delta| \gg \Delta_f$.

Mechanisms other than Raman scattering can cause decoherence to a
hyperfine-state superposition in the presence of light. One such
mechanism is a difference in the Rayleigh scattering rates between
the two levels. A detectable difference in the scattering rate
would identify which state scattered the photons, implying
decoherence. This effect was estimated to be negligible for our
experimental parameters.

In summary, we have measured the decoherence of superposition
states of atomic hyperfine levels caused by spontaneous scattering
of light. The data demonstrate that decoherence is dominated by
Raman inelastic scattering. This rate can be quite small compared
to the total scattering rate if the frequency of the light is
detuned from the relevant excited level by more than its
fine-structure splitting. The total scattering rate therefore
gives a pessimistic measure of decoherence \cite{Wineland_London}.
This has important implications for high-resolution spectroscopy
and quantum-state manipulation.

We thank Chris Oates, John Bollinger and Signe Seidelin for
helpful comments on the manuscript. The work is supported by
NSA/ARDA and NIST. Contribution of NIST; not subject to U. S.
copyright.


\begin{references}
\bibitem{Remark1} In Eq. \ref{rate}, we neglect the Zeeman and hyperfine splitting of the ground and excited states.
  The plots in Fig. \ref{fig3} however, are the result of a calculation that takes these splittings into account.
\end{references}
\end{document}